\documentclass[final,5p,times,twocolumn]{elsarticle}
\usepackage[dvipsnames]{xcolor}
\usepackage{amssymb}
\usepackage{mathrsfs}
\usepackage{physics}
\usepackage{subdepth} 
\usepackage{scalerel}
\usepackage{amsthm}
\usepackage{dsfont}
\usepackage{url}
\usepackage[utf8]{inputenc}
\newcommand{\sm}{\kern0.1em}

\newcommand{\tps}{\skew{3}{\tilde}{\psi}}
\newcommand{\tvph}{\skew{3}{\tilde}{\chi}}
\newcommand*{\paral}{\stretchrel*{\parallel}{0}}

\journal{XXX}

\begin{document}

\begin{frontmatter}

\title{Absorbing detectors meet scattering theory}
\author[first]{Will Cavendish} 
\ead{willcavendish@johnbellinstitute.org}
\affiliation[first]{organization={John Bell Institute for the Foundations of Physics},
            city={New York},
            postcode={NY 10003},
            country={United States}}

\author[second]{Siddhant Das}
\ead{Siddhant.Das@physik.uni-muenchen.de}
\affiliation[second]{organization={Fakultät für Physik, Ludwig-Maximilians-Universität München},
            addressline={Theresienstr.\ 37}, 
            postcode={D-80333 München},
            country={Germany}}

\begin{abstract}
Any proposed solution to the ``screen problem'' in quantum mechanics—the challenge of predicting the joint distribution of particle arrival times and impact positions—must align with the extensive data obtained from scattering experiments. In this paper, we conduct a direct consistency check of the Absorbing Boundary Condition (ABC) proposal, a prominent approach to address the screen problem, against the predictions derived from scattering theory (ST). Through a series of exactly solvable one- and two-dimensional examples, we demonstrate that the ABC proposal's predictions are in tension with the well-established results of ST. Specifically, it predicts sharp momentum- and screen-orientation-dependent detection probabilities, along with secondary reflections that contradict existing experimental data. We conclude that while it remains possible that physical detectors described by the ABC proposal could be found in the future, the proposal is empirically inadequate as a general solution to the screen problem, as it is inconsistent with the behavior of detectors in standard experimental settings.
\end{abstract}


\end{frontmatter}




%
%
%
\section{Introduction}\label{intro}
The study of quantum scattering began with M.\ Born at the dawn of quantum mechanics \cite{Born1926} and has since developed into a core component of modern physics, backed by substantial experimental evidence. Scattering experiments inform our understanding across a wide range of fields, from subatomic particles to solid-state physics. The fundamentals of scattering theory (ST) \cite{Mott,NewtonScattering,RodbergThaler,Amreinscattering} therefore serve as a cornerstone of quantum mechanics. Reproducing the established predictions of ST is a prerequisite for any new theoretical proposal.

In the setting of single-particle potential scattering, the ``scattering into cones'' formula
\begin{equation}\label{PST}
    P_{\text{ST}}\big(\mathcal{D},\psi_0\big) =  \int_{C_{\omega(\mathcal{D})}}\!\!\!\!\! d^3k~\left|\sm\braket{\vb{k}\sm}{\sm\Omega^{\dagger}_-\sm\psi_0}\sm\right|^2
\end{equation}
is a central result in ST. It gives the probability \(P_{\text{ST}}\) of detecting a particle in some detector segment \(\mathcal{D}\), as a function of the initial state \(|\sm\psi_0\rangle\) and the solid angle \(\omega(\mathcal{D})\) subtended by \(\mathcal{D}\) relative to the scattering center. Here, \(C_\omega\) is the cone spanned by \(\omega\), \(\ket{\vb{k}}\) represents the momentum eigenstate of momentum \(\hbar\sm\vb{k}\), and \(\Omega_-\) is the (M{\o}ller) wave operator \cite{Dollard,Combes}. The derivation of the differential scattering cross section from \eqref{PST} is a standard result presented in texts such as \cite[Ch.\ 8]{NewtonScattering} and \cite[p.\ 254]{Weinberg}.

Without scattering potentials, the integrand simplifies to \(|\langle \vb{k} \sm|\sm \psi_0 \rangle|^2\), and the detection probability directly corresponds to the likelihood of finding the particle's momentum within \(C_{\omega(\mathcal{D})}\). Such results enable the determination of momentum distributions from scattering data: e.g., those corresponding to the hydrogen ground state \cite{Lohmann,e2eReview}.

The primary focus of ST is on detection probabilities and scattering cross sections, which provide insights into a wide range of physical processes. However, ST offers an incomplete description of the scattering \emph{process} by failing to account for the statistics of the arrival (detection, impact, flight, or transit) times of the scattered particles.\footnote{Certain characteristic time scales, such as the Eisenbud-Wigner delay and phase times, have been discussed within the ST framework \cite{Muga2008,Bianchi2012}, but their relationship with measured arrival or flight times remains unclear.} 

Our inability to make quantitative predictions about experimental data that we can and do collect in the laboratory constitutes an \emph{empirical} shortcoming of quantum theory. Indeed, many experiments involving screens report both the timing and location of individual detection events; see, e.g., \cite{Pfau, Hediffraction, Ullrich}. However, unlike in textbook measurement theory, the timing of the ``measurements'' involved is not chosen by the experimenter, and unlike in ST, the detection-time is not ``\(\smash{t=\infty}\).'' It is therefore an urgent problem---what Mielnik called the ``screen problem'' \cite[Sec.\ 4]{Bogdan}---to develop an extension of the standard quantum formalism that can predict all observed data.

To fully account for such data, one must be able to describe the joint statistical distribution of impact positions and arrival times of the particle: \(\Lambda\big(\vb{r}, t; \psi_0\big)\), where \(\Lambda\big(\vb{r}, t; \psi_0\big) \, ds \, dt\) gives the probability of detecting a particle in a surface element \(ds\) at position \(\vb{r}\) during the time interval \(dt\) at time \(t\), for some specified initial state \(|\sm\psi_0\rangle\).

Such a joint distribution would solve two crucial problems. First, integrating \(\Lambda\) w.r.t.\ \(\smash{\vb{r}}\) over some detector segment \(\mathcal{D}\) produces the arrival-time distribution as a marginal, thereby solving the notorious ``arrival-time problem'' \cite{MUGA1}. Second, integrating out the detection-time variable from the result, one obtains the probability of detection within \(\mathcal{D}\), \emph{viz.},
\begin{equation}\label{Pw}
    P(\mathcal{D},\psi_0) = \int_0^{\infty}\!\!\!dt\!\int_{\mathcal{D}} ds~\Lambda\big(\vb{r},t;\psi_0\big)\sm,
\end{equation}
where \(ds\) is the surface measure at \(\vb{r}\).

To be physically viable as a general solution to the screen problem, any proposed \(\Lambda\) must have the property that \eqref{Pw} reproduces \(P_{\text{ST}}\) in the regime where ST is applicable and has been experimentally well confirmed, i.e., when the scattering potentials are short-range and the distance between the detection surface and the scattering center approaches infinity.

One prominent proposal to solve the screen problem, developed by Werner and Tumulka \cite{Werner,ABC}, \cite[Sec.\ 5.2]{TumulkaBook}, is based on the idea that the physical detector should influence the particle's dynamics before detection (in contrast to the assumptions underlying Eq.\ \eqref{PST}, where such influences are assumed to be negligible). Their approach incorporates this influence phenomenologically by imposing an ``absorbing boundary condition'' (ABC) on the solutions \(\psi_t\) to the Schr\"odinger equation:
\begin{equation}\label{ABC}
\vu{n}\cdot\pmb{\nabla}\psi_t = \beta \sm \psi_t,
\end{equation}
where \(\vu{n}\) is the outward unit normal vector to the detection surface \(\mathcal{S}\), and \(\beta\) is a complex parameter with \(\smash{\Im\beta>0}\). The joint distribution of detected positions and times is then defined as follows:
\begin{equation}\label{LambdaABC}
\Lambda_{\text{ABC}}\big(\vb{r},t;\psi_0\big) = \frac{\hbar}{m}\sm \Im\beta~|\sm\psi_t(\vb{r})\sm|^2.
\end{equation}

Due to the nature of the imposed boundary condition, the evolution of \(\psi_t\) is \emph{nonunitary}. Specifically, the norm \(\|\sm\psi_t\|\) decreases monotonically over time. The loss of norm is a central feature: \(\|\sm\psi_t\|^2\) represents the probability of non-arrival until time \(t\), making \(\smash{1-\|\sm\psi_t\|^2}\) the cumulative arrival-time probability. This characteristic is common to many approaches to the arrival-time problem that involve explicit detector models, such as \cite{Allcock3,HalliwellKED,Ruschhaupt,Mugaphoton,Hdetect}, and non-unitarity is generally regarded as a consequence of the particle not being isolated when interacting with a physical detector.

Various generalizations of the ABC have been proposed to accommodate particles with nonzero spin \cite{ABCVI, Shadi2024}, as well as for multiparticle scenarios \cite{ABCI}. References \cite{ABCIV,Werner} address the mathematical well-posedness of Schr\"odinger evolution subject to ABCs, while \cite{ABCII,ABCIII,frolov2025,DharEtAl} derive \eqref{ABC} from several different theoretical starting points. (Arrival-) time-energy uncertainty relations implied by \(\Lambda_{\text{ABC}}\) have been explored in \cite{ABCV}. (See also \cite{Kiukas2012}.)

However, except for \cite{aliJavad,AliDDSlit}, relatively little research has focused on the quantitative predictions of the ABC proposal, especially regarding the detection probabilities implied by \eqref{LambdaABC}. This paper takes initial steps to address this gap by comparing these probabilities with \(P_{\text{ST}}\) in the appropriate regime, thus evaluating their consistency with existing experimental results.

In the following section, we present such comparisons through various one- and two-dimensional scattering problems. Our conclusions are summarized in Sec.\ \ref{conclusion}, which also includes a broader discussion.
\section{ABC–ST contrast in the scattering limit}\label{examples}
\subsection{One-dimensional scattering: General considerations}
To begin, consider a simple scattering problem in which a spin-\(0\) particle of mass \(m\), described by the quantum state \(|\sm\psi_0\rangle\) at time zero, moves freely toward a detector located at \(\smash{x=L}\). What is the probability of detection?

As \(L\) approaches infinity, the answer provided by ST [cf.\ Eq.\ \eqref{PST}] is\footnote{From here on, we adopt units where masses, lengths, and times are expressed in units of \(m\), \(\sigma\), and \(m\sm \sigma^2/\hbar\), respectively, where \(\sigma\) is a characteristic length scale of the problem, e.g., the width of \(\psi_0(x)=\expval{x\sm|\sm\psi_0}\).}
\begin{equation}\label{PST1D}
    P_{\text{ST}}\big(\psi_0\big) = \int_0^{\infty}\!\!dk~|\sm\tps_0(k)\sm|^2,
\end{equation}
where \(\tps_0(k)\) is the momentum-space wave function \(\smash{\expval{k\sm|\sm\psi_0}}\). Note that this prediction disregards any specific details related to the detector.

On the other hand, if the detector at \(\smash{x=L}\) behaves according to the ABC proposal \cite{ABC,Werner,DharEtAl}, and assuming that the initial position-space wave function \(\smash{\psi_0(x)=\expval{x\sm|\sm\psi_0}}\) is supported on \((-\infty,L]\), we arrive at the following answer:
\begin{equation}\label{PABC}
    P_{\text{ABC}}\big(\psi_0,L\big) = \Im\beta\!\int_0^{\infty}\!\!\!dt~|\sm\psi_t(L)\sm|^2,
\end{equation}
where \(\psi_t\) solves Schr\"odinger's equation 
\begin{equation}\label{briefsch}
    2\sm i\sm \partial_t\psi_t(x)=-\,\partial_x^2\psi_t(x)\quad \text{for}\quad (x,t)\in (-\infty,L]\times \mathbb{R}_{\ge0}\sm,
\end{equation}
with initial condition \(\psi_0(x)\) and boundary conditions 
\begin{equation}\label{ABCRHS}
    \lim_{x\to-\sm\infty}\psi_t(x)=0,\qquad \text{and} \qquad\partial_x\psi_t(L)=\beta\,\psi_t(L)\sm.
\end{equation}

In \cite{ABC}, the parameter \(\smash{\beta}\) is primarily assumed to be purely imaginary, while \cite{Werner} and \cite{DharEtAl} allow \(\smash{\Re\beta\neq0}\). However, in all cases, it is required that \(\smash{\Im\beta>0}\) to prevent the growth of \(|\sm\psi_t(L)\sm|^2\) over time, and ensure that \eqref{PABC} is nonnegative---both conditions are essential for maintaining the intended probabilistic interpretation of \(P_{\text{ABC}}\).

To compare these two predictions, we introduce the \emph{contrast}
\begin{equation}\label{contrastdef}
    \mathscr{C}_L\big(\psi_0\big) := P_{\text{ST}}\big(\psi_0\big) - P_{\text{ABC}}\big(\psi_0,L\big)\sm,
\end{equation}
which is well defined for all wave functions supported on \((-\infty,L]\). In the scattering limit, as \(\smash{L\to\infty}\), the contrast extends to a function \(\mathscr{C}_{\infty}\) defined on \(L^2(\mathbb{R})\) that can be expressed directly in terms of the momentum-space wave function \(\tps_0\), as follows (see \ref{IT} for details):
\begin{equation}\label{contrast}
    \mathscr{C}_{\infty}\big(\psi_0\big) =\!\int_0^{\infty}\!\!\!dk~|\sm\rho_\beta(k)\sm|^2\sm|\sm\tps_0(k)\sm|^2\!,
\end{equation}
where
\begin{equation}\label{rho}
   \rho_\beta(k)=\frac{k+i\sm\beta}{k-i\sm\beta}.
\end{equation}

It follows that \(\mathscr{C}_{\infty}\) is nonnegative and can only vanish for initial states comprised exclusively of negative momentum components. However, for a generic \(|\sm\psi_0\rangle\), ABC detection probabilities are \emph{lower} than those predicted by ST, irrespective of \(\beta\).\footnote{See, however, Sec.\ \ref{2d} for two-dimensional examples where \(\smash{P_{\text{ABC}}>P_{\text{ST}}}\).}

In order for the ABC proposal to recover the predictions of ST, \(\mathscr{C}_\infty\big(\psi_0\big)\) must be negligible for all experimentally accessible states \(\smash{|\sm\psi_0\rangle\in L^2(\mathbb{R})}\). Equation \eqref{contrast} clearly shows that this does not happen in general, even for a fixed \(|\sm\psi_0\rangle\). 

The \(\beta\)-dependence of \(\mathscr{C}_{\infty}\big(\psi_0\big)\) warrants a detailed examination. Since
\begin{align*}
   |\sm\rho_{\beta}(k)\sm|^2 &= 1 \sm-\sm \frac{4\sm k \Im\beta}{(\Re\beta)^2+(k+\Im\beta)^2}\nonumber\\[2pt]
   &\ge 1 \sm-\sm \frac{4\sm k \Im\beta}{(k+\Im\beta)^2}=\left(\frac{k-\Im\beta}{k+\Im\beta}\right)^2\!,
\end{align*}
we have 
\begin{equation}
    \mathscr{C}_\infty\ge \mathscr{C}_\infty\sm\big|_{\Re\beta=0}
\end{equation}
for any fixed \(|\sm\psi_0\rangle\). That is, purely imaginary \(\beta\)s, as Tumulka \cite{ABC} primarily considers, lower the contrast, making them the best candidates for agreement with ST.

Another aspect of the \(\beta\)-dependence of the ABC that is evident from Eq.\ \eqref{contrast} is that a significant contrast is achieved whenever \(|\sm\beta\sm|\) approaches zero or infinity. This is because in these limits \(\smash{\rho_\beta(k)\to1}\), consequently \(\smash{\mathscr{C}_\infty\big(\psi_0\big) = P_{\text{ST}}\big(\psi_0\big)}\), and \(\smash{P_{\text{ABC}}\big(\psi_0,\infty\big)=0}\). This result can be understood as a consequence of the ABC, Eq.\ \eqref{ABCRHS}, which simplifies to the Neumann boundary condition \(\big(\partial_x \psi_t(L) = 0\big)\) and the Dirichlet boundary condition \(\smash{\big(\psi_t(L) = 0\big)}\), when \(|\sm\beta\sm|\) approaches zero and infinity, respectively. In both scenarios, the evolution is unitary,\footnote{In general, whenever \(\Im\beta=0\); see \cite[p.\ 173]{Exner1985}.} and \(P_{\text{ABC}}\big(\psi_0,L\big)\) (Eq.\ \eqref{PABC}) \emph{vanishes} for any \(|\sm\psi_0\rangle\) and \(L\). As a result, the values of \(\beta\) that are promising for recovering ST predictions should be neither too large nor too small. 
\subsection{Scattering of one dimensional wave packets}\label{1d}
As a concrete example, consider the Gaussian wave packet
\begin{equation}\label{GWP}
    \psi_0(x) = e^{i\sm k_0 x}\,G(x),
\end{equation}
\begin{equation}
    G(x) = \pi^{-\,1/4}\,e^{-\,x^2/2}.
\end{equation}
In momentum space, it is concentrated around \(\smash{k=k_0}\), with
\begin{equation}\label{tpsG}
    \tps_0(k) = G(k-k_0).
\end{equation}
For \(\smash{k_0\gtrsim2}\), ST predicts a nearly \(100\%\) chance of detection, given
\begin{equation}\label{PstOne}
    P_{\text{ST}}\big(\psi_0\big)=1-\frac{1}{2}\operatorname{erfc}(k_0)  \approx 1 - \frac{e^{-\,k_0^2}}{2\,k_0\kern-0.15em\sqrt{\pi}};
\end{equation}
with \(\text{erfc}\) being the complementary error function.

\begin{figure}[!ht]
    \centering
    \includegraphics[width=\columnwidth]{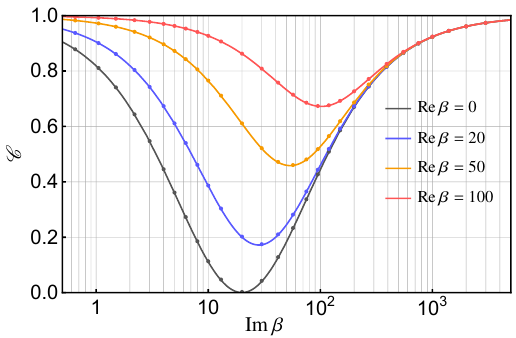}
    \caption{ABC-ST contrast (\(\mathscr{C}_{\infty}\)) curves for the Gaussian wave packet \eqref{GWP} with \(\smash{k_0=20}\) and a fixed \(\Re\beta\) (denoted in the legend). The plot markers indicate \(\mathscr{C}_L\) values for \(\smash{L=2}\), calculated from Eqs.\ \eqref{contrastdef} and \eqref{PABC} using the time-dependent wave packet \(\psi_t^G(x\sm ;k_0,\beta,L)\), Eq.\ \eqref{GSol}.}
    \label{Fig1}
\end{figure}

The value of \(\mathscr{C}_\infty\big(\psi_0\big)\) can be determined using Eq.\ \eqref{contrast}. It is plotted in Fig.\ \ref{Fig1} as a function of \(\Im\beta\) for select values of \(\Re\beta\). The plot reveals several key features of the proposal's \(\beta\)-dependence: the contrast exhibits symmetry w.r.t.\ the sign of \(\Re\beta\) (so it suffices to consider only \(\smash{\Re\beta \ge 0}\)); it reaches its minimum when the value of \(\beta\) is tuned to the central momentum of the wave packet, specifically when \(\smash{\beta \approx i\sm k_0}\); and the contrast approaches \(\smash{P_{\text{ST}} \approx 1}\) for very small and large values of \(|\sm\beta\sm|\), as noted earlier.

The initial boundary value problem (IBVP) defined by Eqs.\ (\ref{briefsch}--\ref{ABCRHS}) and \eqref{GWP} is analytically solvable (see \ref{task2} for details) for any \(L\). The solution, denoted as \(\psi_t^G\) for future reference, takes the form:
\begin{align}\label{GSol}
    \psi^G_t\big(x\sm ;k_0,\beta,L\big) &= \varphi_t^G\big(x\sm ;k_0\big) \,+\, \varphi_t^G\big(2\sm L-x\sm ;k_0\big) \,+\,   \big(2\sqrt{\pi}\,\big)^{1/2}\sm \beta\sm\nonumber \\
    &\kern5mm  \times\sm \exp(-\,\frac{1}{2}\sm\big(k_0-i\sm\beta\big)^2+\frac{i\sm t}{2}\sm\beta^2-\beta\sm (2\sm L-x)) \nonumber \\
    &\kern4.8mm\times\sm \text{erfc}\!\left(\frac{2\sm L-x-i\sm k_0-\beta\sm(1+i\sm t)}{\sqrt{2}\sqrt{1+i\sm t}}\right)\kern-0.1em,
\end{align}
where
\begin{equation}\label{psitG}
    \varphi_t^G\big(x\sm ;k_0\big) = \frac{e^{-\,k_0^2/2}}{\sqrt{1+i\sm t}}\,G\kern-0.1em\left(\frac{x-i\sm k_0}{\sqrt{1+i\sm t}}\right)
\end{equation}
solves the free-particle Schr\"odinger equation for the same initial condition and usual boundary conditions, \emph{viz.}, \(\smash{\varphi_t^G(x)\to0}\), as \(\smash{|x|\to\infty}\).

By substituting this result into Eq.\ \eqref{PABC}, we can numerically evaluate \(P_{\text{ABC}}\big(\psi_0,L\big)\) for any finite \(L\). This provides a direct way to assess how rapidly \(\mathscr{C}_L\) approaches its far-field limit, Eq.\ \eqref{contrast}. As shown by the plot markers in Fig.\ \ref{Fig1}, even \(L=2\) yields near-perfect agreement between \(\mathscr{C}_L\) and \(\mathscr{C}_{\infty}\). This demonstrates that the ABC detection probabilities quickly converge as \(L\) increases, and even in the \emph{near-field regime}, they align perfectly with the asymptotic predictions.

The current example suggests that the parameter \(\beta\) characterizing an ABC detector could be tuned to match the specified \(|\sm\psi_0\rangle\) in order to recover ST predictions. In other words, an ABC detector could be effective in detecting only specific wave functions and no others, in a manner consistent with ST and experiment. Although this idea may seem reasonable at first glance, it raises several questions if taken as the intended interpretation of the ABC proposal. We discuss these one by one in what follows.

\paragraph{(1) High energy scattering} The ABC proposal leads to a surprising prediction: particles with sufficiently \emph{high} momenta \((\gg \Im \beta)\) are likely to be backscattered by the detector without leaving any detectable trace. This prediction poses a significant challenge to the ABC proposal in light of the demonstrated success of ST in high-energy experiments.

\paragraph{(2) POVM structure} Introducing a \(|\sm\psi_0\rangle\)-dependent \(\beta\), i.e., allowing \(\smash{\beta = \beta\big(\psi_0\big)}\), would compromise the \emph{bilinearity} of \(\Lambda_{\text{ABC}}\), Eq.\ \eqref{LambdaABC}, w.r.t.\ \(|\sm\psi_0\rangle\). For theorists who consider the Positive Operator Valued Measure (POVM) framework \cite{Busch1995}, \cite[Ch.\ 5]{TumulkaBook} to be foundational, such a modification would be unacceptable.\footnote{It is worth noting that \(P_{\text{ST}}\), Eq.\ \eqref{PST}, corresponds to a POVM, which significantly differs from those resulting from the ABC proposal.} On this view, the parameter \(\beta\) must be the same for \emph{all} admissible wave functions.

\paragraph{(3) Coherent superpositions and statistical mixtures} Even if \(\beta\) were allowed to vary according to the wave function, there exist numerous wave functions for which no selection of \(\beta\)---purely imaginary or otherwise---would result in alignment with ST. Consider, for instance, the superposition of Gaussian wave packets
\begin{equation}\label{Gsup}
    \psi_0(x) = \frac{G(x)}{\sqrt{2\sm N}}\,\Big(e^{i\sm k_0 x} + e^{i\sm k_1 x}\Big)\sm,
\end{equation}
where \[\smash{N = 1 + \exp(-\,(k_0 - k_1)^2\kern-0.1em / 4)}\] is the required normalization constant. When exposed to an ABC detector at \(L\), it evolves into
\begin{equation}\label{Phit}
    \psi_t(x) = \frac{1}{\sqrt{2\sm N}}\,\Big(\psi_t^G\!\left(x\sm ;k_0,\beta,L\right) \,+\, \psi_t^G\!\left(x\sm ;k_1,\beta,L\right)\kern-0.1em\Big)\sm,
\end{equation}
since the ABC-evolution is linear. 

\begin{figure}[!ht]
    \centering
    \includegraphics[width=\columnwidth]{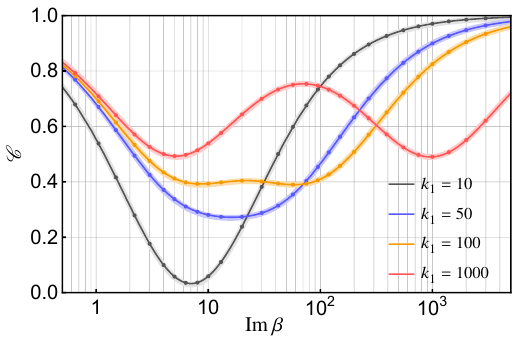}
    \caption{ABC-ST contrast (\(\mathscr{C}_{\infty}\)) curves for the superposition of Gaussian wave packets Eq.\ \eqref{Gsup}, with \(\smash{\Re\beta = 0}\), \(\smash{k_0 = 5}\), and a fixed \(k_1\) (denoted in the legend). Plot markers indicate \(\mathscr{C}_L\) values for \(\smash{L = 10}\), calculated using Eqs.\ \eqref{contrastdef}, \eqref{PABC}, and \eqref{Phit}. Thick curves depict contrasts calculated from Eq.\ \eqref{Capprox}.}
    \label{Fig2}
\end{figure}

In this case, it turns out that neither \(\smash{\beta \approx i\sm k_0}\) nor \(i\sm k_1\) provides close agreement between ST and ABC detection probabilities once \(k_1\) exceeds \(k_0\) by even a modest margin. This is evident even in the optimal situation when \(\smash{\Re \beta = 0}\), illustrated in Fig.\ \ref{Fig2}. In particular, the contrast becomes uniformly bounded from below; therefore, it cannot be further lowered by fine-tuning \(\beta\). 

This uniform lower bound can be explained by a simple approximation of Eq.\ \eqref{contrast} by the Laplace method, which produces
\begin{equation}\label{Capprox}
    \mathscr{C}_{\infty}\big(\psi_0\big) \approx \frac{|\sm\rho_\beta(k_0)\sm|^2 + |\sm\rho_\beta(k_1)\sm|^2}{2\sm N}.
\end{equation}
The right-hand side attains a global minimum value of \[\left(1-4\sm k_0\sm k_1/(k_0-k_1)^2\right)/(2\sm N),\] which approaches \(0.5\) in the limit \(\smash{|k_1-k_0|\to\infty}\); this is also indicated by the \(\smash{k_1=1000}\) curve in Fig.\ \ref{Fig2}. In summary, it is not always possible, even in theory, to have a value of \(\beta\) that guarantees agreement between ST and the ABC proposal for a given generic wave function; contrary to what the earlier Gaussian wave packet example may have led us to expect.

A similar argument applies to statistical mixtures (as opposed to coherent superpositions) of wave packets with different momenta. In any experimental scenario that involves a broad range of incoming momenta---such as in the double-slit experiment \cite{Pfau,wig}, where metastable He atoms were sourced from a gas-discharge tube over a broad range of velocities \cite[Sec.\ 3]{wig}---the ABC proposal imposes stringent upper bounds on the overall detection efficiency, regardless of the spatial separation or timing of the incoming components.

\paragraph{(4) Real-world detectors} Do real-world detectors actually exhibit the peculiar momentum sensitivity implied by the ABC proposal?  While detector design details can vary, many scattering experiments report nearly constant efficiency across a wide range of energies.  For example, Lohmann and Wiegold \cite{Lohmann} demonstrate excellent agreement between measured data and the theoretical momentum distribution over a wide range of incoming momenta, indicating that the detector sensitivity is essentially momentum-independent within the experimental precision.

\begin{figure}[!ht]
        \centering 
        \includegraphics[width=\columnwidth]{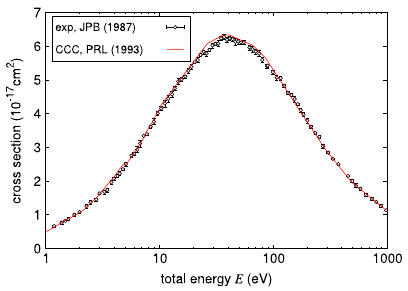}
        \caption{Total scattering cross section for ionization of hydrogen atoms by electron impact. Excess energy\protect\footnotemark above \(13.6 \text{ eV}\) versus scattering cross-section (adapted from Bray \emph{et al.} \cite{bray2012electron} with permission from the publisher), showing excellent agreement between ST and experiment (see text for additional details).}
        \label{Fig3}
\end{figure}

The experimental data from Shah \emph{et al.} \cite{Exp1987} further reinforce this point. They report total cross-section measurements for the same electron-hydrogen-atom collision experiment, with incoming electron energies ranging between \(\smash{14\text{ eV}}\) and \(\smash{4000\text{ eV}}\). This data was plotted by Bray \emph{et al}.\ \cite{bray2012electron} to demonstrate the close agreement between the convergent close-coupling (CCC) method for computing scattering cross sections and Shah \emph{et al.}'s measurements (see Fig.\ \ref{Fig3} below). It should be emphasized that the CCC method is an ab initio numerical method for calculating the predictions of ST for many-body Coulomb Hamiltonians with no adjustable parameters. Therefore, the close agreement exhibited in Fig.\ 3 is evidence for the empirical success of ST.

Importantly, Shah \emph{et al.} reported that the overall efficiency of ion detection in their setup ``was found to remain constant over our full energy range'' \cite[p.\ 3507]{Exp1987}. This experimental result is therefore in direct conflict with the sharp momentum-dependent detection efficiency predicted by the ABC proposal.
\subsection{Scattering of two dimensional wave packets}\label{2d}
\footnotetext{Although the horizontal-axis is labeled ``total energy'' in \cite{bray2012electron}, the plotted values actually correspond to the total energy data from \cite{Exp1987} with \(13.6 \, \text{eV}\) subtracted.}
While this momentum-dependent efficiency is observed in any number of spatial dimensions, higher-dimensional scenarios reveal further conflicts between ST and the ABC proposal. As we shall see in this subsection, these conflicts arise from the dependence of detection probabilities on the overall geometry and orientation of the detection surface.

\begin{figure}[!ht]
    \centering
    \includegraphics[width=\columnwidth]{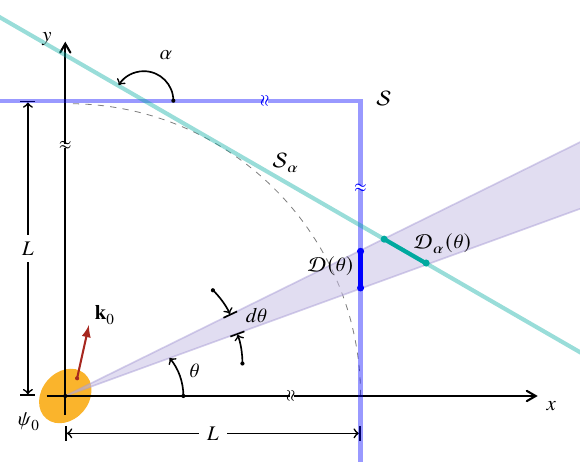}
    \caption{A wave packet \(\psi_0\) with mean momentum \(\vb{k}_0\) approaches either an L-shaped screen \(\mathcal{S}\) or a flat screen \(\mathcal{S}_\alpha\) tilted at angle \(\alpha\) relative to the \(x\)-axis. The infinitesimal detector segments \(\smash{\mathcal{D}(\theta)\subset\mathcal{S}}\) and \(\smash{\mathcal{D}_{\alpha}(\theta)\subset\mathcal{S}_{\alpha}}\) both span the same convex cone of apex angle \(d\theta\) (shaded area).}
    \label{Fig4}
\end{figure}

To illustrate these conflicts, we examine two scattering scenarios shown in Fig.\ \ref{Fig4}, which differ only in the shape of the detection surface. In one scenario, we have a flat screen \(\mathcal{S}_\alpha\) inclined at an angle \(\alpha\) w.r.t.\ to the \(x\)-axis. In the second scenario, we take an L-shaped screen \(\mathcal{S}\). In both cases, the particle is represented by the two-dimensional Gaussian wave packet
\begin{equation}\label{IC2D}
    \psi_0(\vb{r}) = G\big(x\big)\,G\big(y\big)\,e^{i\sm \vb{k}_0\cdot \vb{r}}\kern-0.1em,
\end{equation}
at \(\smash{t=0}\). As time progresses, the wave function propagates toward the detection surface. Our goal is to determine the probability of detecting the particle in the infinitesimal detector segment \(\mathcal{D}_\alpha(\theta)\) or \(\mathcal{D}(\theta)\) depicted in Fig.\ \ref{Fig4}.\footnote{For these segments to exist, the angle \(\theta\) must be restricted to the intervals \(\smash{(\alpha-\pi,\alpha)}\) and \(\smash{(-\sm\pi/2,\pi)}\) for the planar and L-shaped screens, respectively.}

First, for the inclined screen 
\begin{equation}
    \mathcal{S}_\alpha=\left\{\sm \vb{r}\in\mathbb{R}^2\kern-0.1em: \vu{n}(\alpha)\cdot\vb{r}=L\sm\right\}\kern-0.1em,
\end{equation}
with outward unit normal vector
\begin{equation}
    \vu{n}(\alpha)=\sin\alpha\,\vu{e}_x-\cos\alpha\,\vu{e}_y\sm,
\end{equation}
the (differential) detection probability predicted by ST in the limit as \(\smash{L\to\infty}\), is 
\begin{align}\label{STerf}
    dP_{\text{ST}}(\theta) &\overset{\eqref{PST}}{=} d\theta\!\int_0^{\infty}\!\!dk~k\,|\langle\vb{k}\sm|\sm\psi_0\rangle|^2 \nonumber\\
    &= d\theta~ G^2\big(k_0\big)\!\int_0^{\infty}\!\!dk~k~G^2\big(k\big)\sm\exp\kern-0.1em\Big(2\sm k\sm k_0\cos\kern-0.1em\left(\theta-\theta_0\right)\!\Big)\sm,
\end{align}
where \(\smash{\theta_0 = \text{Arg}\sm\big(k_{0x}+i \sm k_{0y}\big)}\). Applying Laplace's method, we conclude that the integral is concentrated around \(\smash{\theta = \theta_0}\) (see Fig.\ \ref{Fig7} below).

Note that \eqref{STerf} does not depend on \(\alpha\). This highlights a key characteristic of \(P_{\text{ST}}\): it is determined solely by the convex cone formed by the detector segment in question. The overall shape or orientation of the detection surface to which the segment belongs does not affect this probability. Specifically, segments from different detection surfaces that cover the same angle relative to the scattering center (e.g., \(\smash{\mathcal{D}_\alpha(\theta) \subset \mathcal{S}_\alpha}\) and \(\smash{\mathcal{D}(\theta) \subset \mathcal{S}}\)) will have an equal likelihood of detecting the particle \emph{in the scattering limit}.

However, the ABC detection probability does not exhibit this property. To illustrate this, note that the differential probability for detection in \(\smash{\mathcal{D}_\alpha}\), according to the ABC proposal, is
\begin{equation}\label{dPABC}
    dP_{\text{ABC}}(\theta) =  d\ell\,\Im\beta\!\int_0^{\infty}\!\!\!dt\,\left|\sm\psi_t\big(\vb{R}_{\theta}\big)\sm\right|^2\!,
\end{equation}
where \[\vb{R}_{\theta} = L\csc(\alpha-\theta)\sm\Big(\kern-0.1em\cos\theta\,\vu{e}_x+\sin\theta\,\vu{e}_y\Big)\sm,\] and \(\smash{d\ell= L\csc^2(\alpha-\theta)\sm d\theta}\) is the line element or the length of the segment \(\mathcal{D}_\alpha(\theta)\). The wave function \(\psi_t\) satisfies the free Schr\"odinger equation 
\begin{equation}\label{sch2d}
    2\sm i\sm \partial_t\sm \psi_t =-\laplacian\psi_t
\end{equation}
in the half-spaces bounded by \(\mathcal{S}_\alpha\) containing the support of \(\psi_0\), and the ABC \eqref{ABC}, with \(\vu{n}=\vu{n}(\alpha)\), and initial condition \eqref{IC2D}.

Using the results from the previous section and standard separation of variables techniques, we can determine the wave function that solves this IBVP, obtaining
\begin{equation}\label{phi2D}
    \psi_t(\vb{r}) = \psi^G_t\big(r_{\paral}\sm;k_{0\paral},\beta,L\big)\, \varphi^G_t\big(r_{\!\perp};k_{0\perp}\big)\sm,
\end{equation}
where
\begin{equation}
  r_{\paral} = \vu{n}\big(\alpha\big)\cdot \vb{r},\quad  r_{\!\perp} = \vu{n}\big(\alpha+\pi/2\big)\cdot \vb{r},
\end{equation}
and \(k_{0\paral}\) and \(k_{0\perp}\) are defined analogously. In these coordinates, the wave equation \eqref{sch2d}, the initial wave function \eqref{IC2D}, and the relevant ABC all separate. The evolution parallel to the screen is free, while the evolution perpendicular to the screen follows the one-dimensional IBVP solved in Subsec.\ \ref{1d}.

\begin{figure}[!ht]
    \centering
    \includegraphics[width=\columnwidth]{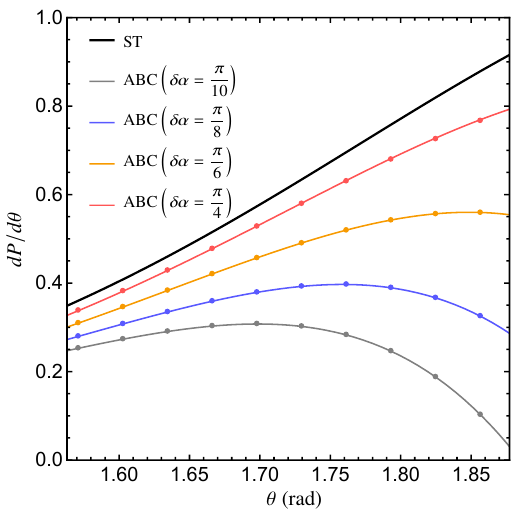}
    \caption{Angular variation of the differential detection probability for a two-dimensional Gaussian wave packet of central momentum \(\smash{\vb{k}_0 = \sqrt{3}\,\vu{e}_y - \vu{e}_x}\) approaching the inclined screen \(\mathcal{S}_\alpha\). The black curve shows the \(\alpha\)-independent ST prediction, while colored curves represent predictions of the ABC proposal for \(\smash{\beta = 2 \sm i}\) and different tilt angles \(\smash{\alpha = \pi/2 + \delta\alpha}\) (see legend). Plot markers indicate numerical values derived from Eq.\ \(\eqref{dPABC}\) for a finite detector distance of \(\smash{L = 15}\).}
    \label{Fig5}
\end{figure}

Finally, incorporating \eqref{phi2D} into \eqref{dPABC} and letting \(\smash{L\to\infty}\), we arrive at the exact result:
\begin{align}
    \frac{dP_{\text{ABC}}}{d\theta} &= 4\sm G^2(k_0)\sm\Im\bar{\beta}\!\int_0^{\infty}\!\!\frac{dk}{|k-i\sm\bar{\beta}\sm|^2} \sm k^2\sm G^2(k)\nonumber\\
    &\kern3cm\times\exp\kern-0.1em\Big(2\sm k\sm k_0\cos\kern-0.1em\left(\theta-\theta_0\right)\!\Big)\sm.
\end{align}
Here,
\begin{equation}
    \bar{\beta} = \beta\csc(\alpha-\theta),
\end{equation}
through which the result explicitly depends on \(\alpha\), unlike \eqref{STerf}. This \(\alpha\)-dependence is illustrated in Fig.\ \ref{Fig5}. As our formulas show and the figure confirms, the ABC prediction disagrees with ST for any choice of the screen inclination angle \(\alpha\).

This \(\alpha\)-dependence has implications that extend beyond the simplified scenario of a single freely moving particle. For instance, in the context of elastic scattering of multiple quantum particles interacting via a central potential, the Hamiltonian's rotational symmetry implies, according to ST, that the scattering cross sections depend solely on the \emph{relative scattering angle} between the incoming and outgoing particles \cite[Ch.\ 10]{RodbergThaler}. An ideal detection model must respect this symmetry. However, the ABC model---particularly its generalization suitable for multiparticle problems \cite{ABCI}---does not. Specifically, it predicts unexpected orientation-dependent cross sections, suggesting that the scattering pattern should vary depending on the alignment of the colliding particle beams relative to the laboratory, contradicting both ST and experimental observations.

We will now discuss the case of the L-shaped screen
\begin{equation}
    \mathcal{S} = \left\{\sm\mathbf{r} \in \mathbb{R}^2 : \max(x,y) = L\sm \right\}.
\end{equation}
The ST prediction for the probability of detection along its segment \(\mathcal{D}(\theta)\), shown in Fig.\ \ref{Fig4}, is the same as \eqref{STerf}.

\begin{figure}[!ht]
    \centering
    \includegraphics[width=\columnwidth]{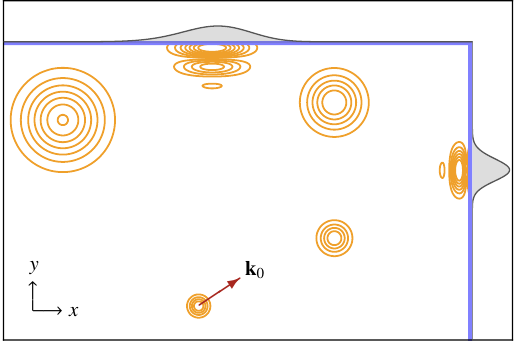}
    \caption{A selection of contour plots illustrating the development of \eqref{exactsol} over time. A wave packet approaching an L-shaped ABC-screen generates screen-position probability distribution functions with two distinct peaks: the first peak, which would also be present for a flat screen, and the second peak, resulting from the portion of the wave packet that is reflected.}
    \label{Fig6}
\end{figure}

To determine the corresponding ABC prediction, we need the solution \(\psi_t\) of Eq.\ \eqref{sch2d} defined on \(\big\{\vb{r}\in\mathbb{R}^2:\max(x,y)<L\big\}\times \mathbb{R}_{\ge0}\) and subject to the ABCs [cf.\ Eq.\ \eqref{ABC}]
\begin{align}
    &\partial_x\psi_t\big(L,y\big)=\beta\sm \psi_t\big(L,y\big),\qquad y<L, \nonumber\\[2pt]
    &\partial_y\psi_t\big(x,L\big)=\beta\sm \psi_t\big(x,L\big),\qquad x<L.
\end{align}
For the Gaussian initial condition \eqref{IC2D}, the IBVP is separable as before, allowing us to express the exact solution in the form
\begin{equation}\label{exactsol}
    \psi_t(\vb{r}) =  \psi^G_t\big(x\sm; k_{0x},\beta,L\big)\,\psi^G_t\big(\sm y\sm ;k_{0y},\beta,L\big),
\end{equation}
where \(\psi_t^G\) is defined by \eqref{GSol}. We illustrate the evolution of the wave packet as it approaches the L-shaped screen in Fig.\ \ref{Fig6}. 

Initially, the wave packet interacts with the vertical segment of the screen, where a portion of it is absorbed. This results in a peak in the detection probability distribution along the vertical segment of \(\mathcal{S}\). Meanwhile, a component of the wave packet that is reflected upon impact subsequently contacts the horizontal segment of \(\mathcal{S}\), leading to the formation of a distinct second detection peak---a distinctive prediction of the ABC proposal expected for a wide range of parameters. However, it would be entirely anomalous from the point of view of ST, which predicts only a single peak, as per Eq.\ \eqref{STerf}.

\begin{figure}[!ht]
    \centering
    \includegraphics[width=\columnwidth]{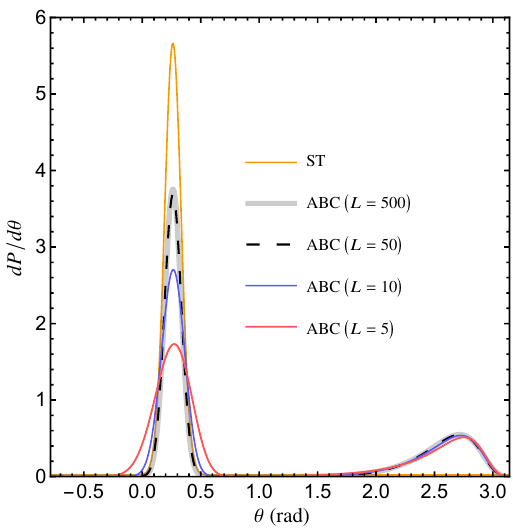}
    \caption{Angular variation of the differential detection probability for a two-dimensional Gaussian wave packet of central momentum \(\smash{\vb{k}_0 = 9.66\,\vu{e}_x + 2.59\,\vu{e}_y}\) approaching the L-shaped surface \(\mathcal{S}\). The colored curves represent the results from the ABC proposal for \(\smash{\beta = 2.59\sm i}\) at various finite detector distances (indicated in the legend). The ABC curves show minimal variation beyond \(\smash{L=50}\), and each one presents two distinct peaks, unlike the single peak predicted by ST (orange curve).}
    \label{Fig7}
\end{figure}

The second peak is not an artifact of near-field scattering, as it remains clearly visible even in the scattering limit \(\smash{L\to\infty}\). To further support this claim, we evaluate the ABC prediction for the probability of detection in \(\mathcal{D}(\theta)\) using Eq.\ \eqref{dPABC} with \(\psi_t\) from \eqref{exactsol}. We set \(\vb{R}_{\theta} = L/\max\kern-0.1em\big(\kern-0.1em\cos\theta,\sin\theta\big)~\big(\kern-0.1em\cos\theta\,\vu{e}_x+\sin\theta\,\vu{e}_y\big)\) and \(\smash{d\ell= L\sm R_{\theta}^{-2}d\theta}\). Figure \ref{Fig7} presents numerical results for increasing \(L\) values, which confirms the convergence of \(dP_{\text{ABC}}/d\theta\) and the persistence of the second peak in the scattering limit.

Note that in this limit, the ABC proposal predicts a \(66\%\) (\(33\%\)) chance of detection in the vertical (horizontal) section of \(\mathcal{S}\). In contrast, ST predicts nearly a \(100\%\) chance of detection only in the vertical section. The L-shaped screen example thus compellingly illustrates why the predictions of the ABC proposal can differ dramatically from those of ST. 

\section{Discussion}\label{conclusion}
Developing a predictive theory of particle arrival times and impact positions---the ``screen problem''---remains a pressing challenge for both theorists and experimentalists. The ABC proposal offers a mathematically compelling contribution to this discussion. However, any widely applicable solution must meet the constraints imposed by more than a century of successful scattering experiments. As we have demonstrated, the ABC proposal is incompatible with scattering data obtained from real-world detectors. While it is conceivable that detectors described by the ABC proposal might be found in the future (see below), existing experimental data seem to rule out the ABC proposal as a general solution to the screen problem.

We found that the proposal leads to unexpected momentum-dependent detection probabilities. This momentum dependence introduces a bias in the measured cross-sections and sets upper limits on detector efficiency, which conflicts with experimental observations. Additionally, the proposal predicts significant secondary peaks due to wave packet components reflecting off the surfaces of the detector. While we are not aware of any dedicated studies involving L-shaped screens, it seems highly improbable that such reflections are a common occurrence. If they were, it would require a reevaluation of the interpretation of countless experiments.

Take, for instance, Rutherford's gold-foil experiment, where the infrequent backscattering of alpha particles---about \(1\) in \(20,000\)---was a vital clue that led to the discovery of the atomic nucleus. If the detectors themselves were responsible for reflections, it would call into question the interpretation of this and countless subsequent scattering experiments. The remarkable success of modern Angle-Resolved Photoemission Spectroscopy (ARPES), which is commonly used to determine the band structure of materials, also presents a direct challenge to the ABC proposal. Note that ARPES relies on the core assumption that photoelectrons, when liberated from a sample, travel in a straight line to the detector. However, the secondary reflections predicted by the ABC proposal would scramble the measured angles, rendering the data uninterpretable.

An intriguing body of work cited in Sec.\ \ref{intro} suggests that the ABC can be derived from more fundamental principles, including the dynamics of the measurement process \cite{frolov2025} and from variants of spontaneous or objective collapse models \cite{ABCII} or from chop-and-evolve schemes encountered in discussions of the quantum Zeno effect \cite[Sec.\ V.\ A]{DharEtAl}. To the extent that these derivations are robust consequences of those foundational assumptions, the empirical shortcomings pointed out in this paper call into question not just the ABC proposal, but also the assumptions themselves.

Despite the conflict with existing experimental data, it is still possible that the ABC proposal gives a good description of physically realizable detectors in certain situations. The key question, then, is what sorts of detectors those might be. The answer is far from obvious. Tumulka's ``ideal detector hypothesis'' \cite{ABC} posits that the ABC proposal parameterizes ``ideal'' detector models. Such models, he suggests, should ``disregard physical details that may vary from one detector to another and would represent the `best possible' detector, to which real detectors should be designed to approximate.'' However, ABC detector models do not seem ``ideal'' in the intended sense. Conventional ST already disregards detector details that vary from one detector to another, and also matches observed data from real (``non-ideal'') detectors. As we have shown, ABC detector models distort the data to a greater extent than the experimental data suggests is necessary.

In several papers: \cite{ABC}, \cite{ABCIII}, and \cite{ABCV}, Tumulka indicates that the proposal should apply to ``hard'' detectors. He explains that a hard detector bounding a region \(\smash{\Omega\subset\mathbb{R}^3}\) ``detects the particle as soon as it reaches \(\partial \Omega\), as opposed to a `soft' detector that may take a while to notice a particle moving through the detector volume'' \cite{ABCV}. However, at present, ``hardness'' and ``softness'' are heuristic notions, lacking the precision needed to determine whether the proposal applies to any physical detector. Therefore, operationalizing the distinction between ``hard'' and ``soft'' detectors is an essential next step for ABC proponents who share this view.

Far from discouraging further study, our critique emphasizes the importance of continuing to investigate the ABC proposal and its various generalizations. Identifying real-world detectors described by an ABC would be a significant discovery, yielding empirical signatures that differ markedly from the predictions of standard ST. If proponents of ABC view these differences as more than mere theoretical artifacts, they should actively encourage experiments designed to detect them.

The shortcomings of the ABC proposal we have pointed out should not overshadow the urgency of the issue it was intended to address. The screen problem remains a central and unresolved challenge in quantum theory. While ST has been remarkably successful within its specific domain, it does not provide a complete account of detection events---specifically, when and where physical detectors register particles. The ABC proposal presents a mathematically elegant approach, but its conflict with experimental data only underscores the need for a model that can handle general detector configurations while also recovering the established predictions of scattering theory.

Crucially, we recommend that further developments in this field be guided by empirical data, including both re-examinations of existing results and the design of new, targeted experiments. Such input is essential if we hope to extend the empirical success of ST into regimes where its underlying assumptions begin to break down.

\section*{Acknowledgements}
The authors thank Lawrence Frolov for helpful discussions on ABCs and for bringing Ref.\ \cite{Exner1985} to their attention. They also thank Roderich Tumulka for sharing his perspective on the role of ``hardness'' in defining the scope of the ABC proposal, and Catalina Curceanu for her detailed feedback and insights regarding various experimental aspects.
\appendix

\section{~}\label{IT}
In this section, we will:
\begin{enumerate}[(i)]
    \item Express the contrast \(\smash{\mathscr{C}_L\big(\psi_0\big)}\) (Sec.\ \ref{1d}) in terms of the momentum-space wave function \(\tps_0(k)\) in the limit as \(\smash{L \to \infty}\). \label{O1}
    \item Derive an analytic expression for \(\psi_t(x)\) corresponding to an initial Gaussian wave packet, Eq.\ \eqref{GWP}. \label{O2}
\end{enumerate}
To achieve this, we need a suitable representation of the function \(\psi_t\), which solves the initial boundary value problem (IBVP) defined by Eqs.\ \eqref{briefsch} and \eqref{ABCRHS}. 

To that end, following \cite{DtoR}, express the solution implicitly as follows:
\begin{equation}\label{ansatz}
    \partial_x\psi_t - \beta \sm \psi_t = \chi_t,
\end{equation}
introducing the function \(\chi_t(x)\) that solves Eq.\ \eqref{briefsch} for \((x,t) \in (-\infty,L] \times \mathbb{R}_{\ge0}\), subject to the boundary conditions:
\begin{equation}\label{BCs}
    \lim_{x\to-\sm\infty}\chi_t(x) = 0 = \chi_t(L),
\end{equation}
and initial condition
\begin{equation}\label{initvphi}
    \chi_0\overset{\eqref{ansatz}}{=}\partial_x\psi_0 - \beta\sm \psi_0.
\end{equation}
It follows that \(\psi_t\) defined by \eqref{ansatz} meets the ABC at \(L\).

Assuming \(\smash{\Re\beta<0}\) for mathematical convenience,\footnote{The case where \(\smash{\Re \beta = 0}\) will be addressed shortly. However, the result of \ref{task2} remains valid even when \(\smash{\Re \beta>0}\).} the general solution of Eq.\ \eqref{ansatz}---a first-order inhomogeneous ODE---takes the form
\begin{equation}\label{sol}
    \psi_t(x) = A\,e^{\sm\beta\sm  x} + \int_{-\infty}^x\!\!dx^\prime~\chi_t\big(x^\prime\big)\,e^{\sm\beta \sm (x\sm -\sm x^\prime)},
\end{equation}
where \(A\) is an arbitrary constant. In view of \eqref{initvphi}, the right-hand side reproduces \(\psi_0\) at time zero iff \(\smash{A=0}\), as can be verified through partial integration. When \(\smash{A=0}\) holds, \eqref{sol} immediately implies that $\psi_t$ satisfies the boundary condition $\displaystyle \lim_{x\to-\infty} \psi_t(x)=0$ in Eq.\ \eqref{ABCRHS} as well.

It remains to demonstrate that the obtained \(\psi_t\) satisfies Schr\"odinger's equation for \(\smash{(x,t)\in (-\infty,L]\times \mathbb{R}_{\ge0}}\). For this, it is helpful to reformulate \eqref{sol} as
\begin{equation}\label{sol1}
    \psi_t(x) = \int_0^{\infty}\!\!dx^\prime~\chi_t\big(x-x^\prime\big)\,e^{\sm \beta \sm x^\prime},
\end{equation}
using a simple substitution. Now, applying \(2\sm i\sm \partial_t + \partial_x^2\) to the right-hand side and noting that \(\chi_t\) itself solves Eq.\ \eqref{briefsch}, one finds that \(\smash{2\sm i\sm \partial_t\psi_t + \partial_x^2\psi_t=0}\).

Next, we evaluate \(\chi_t\). For \(\psi_0\) (and consequently, \(\chi_0\), by Eq.\ \eqref{initvphi}) compactly supported on \((-\infty,L]\), which we assume for convenience,\footnote{As we'll see in \ref{task2} below, the compactness restriction is not essential.} \(\chi_t\) is determined by the so-called ``method of images'' \cite{MOI}.  We express the solution in the form
\begin{equation}\label{MOI}
    \chi_t(x) = \int_{-\infty}^{\infty}\!\frac{dk}{\sqrt{2\sm \pi}}~\tvph_0(k)\, \left(e^{i\sm k\sm x}\, -\, e^{i\sm k\sm (2\sm L-x)}\right)\sm e^{-\,i\sm t\sm k^2/2},
\end{equation}
where
\begin{equation}\label{initvarphik}
    \tvph_0(k) \overset{\eqref{initvphi}}{=} \big(i\sm k-\beta\sm\big)\sm\tps_0(k).
\end{equation}
Evidently, \(\smash{\chi_t(L)=0}\) and the right-hand side satisfies Eq.\ \eqref{briefsch}, correctly reproducing the initial condition \eqref{initvphi} on \((-\infty,L]\), at \(\smash{t=0}\).

Substituting \eqref{MOI} into \eqref{sol1}, we obtain:
\begin{equation}\label{phitx}
    \psi_t(x) = \int_{-\infty}^{\infty}\!\frac{dk}{\sqrt{2\sm \pi}}~\tps_0(k)\sm\left(e^{i\sm k\sm x}\, +\, \rho_\beta(k)\sm e^{i\sm k\sm (2\sm L-x)}\right)\sm e^{-\,i\sm t\sm k^2/2},
\end{equation}
where
\begin{equation}\label{rLk}
    \rho_\beta(k) = \frac{k+i\sm\beta}{k-i\sm\beta}.
\end{equation}
Note that the expression within parentheses satisfies the ABC at \(L\)---a linear boundary condition---clarifying why Eq.\ \eqref{phitx} solves the BVP under consideration.

The case \(\smash{\Re\beta=0}\) (say, \(\smash{\beta = i\sm \kappa}\) with \(\smash{\kappa>0}\)) is particularly relevant for Sec.\ \ref{examples}, but we have not yet addressed it in our derivation. Eq.\ \eqref{phitx} is typically invalid in this scenario because \(\rho_\beta\) is singular at \(\smash{k = -\,\kappa}\). It turns out that \(\psi_t\) in this case is obtained as the limit of Eqs.\ (\ref{phitx}-\ref{rLk}) for \(\smash{\beta = \epsilon+i\sm \kappa}\), as \(\epsilon\) approaches zero from below. For brevity, we will skip the proof that this prescription indeed solves the IBVP in question.

This concludes our analysis of the IBVP concerning \(\psi_t\). We will now proceed to complete tasks \eqref{O1} and \eqref{O2} in the following sections.
\subsection{~}\label{task1}
Note that the function \(\psi_t(x)\) given by Eq.\ \eqref{phitx} is well-defined for any \((x,t)\in \mathbb{R}\times\mathbb{R}_{\ge0}\), even though its physical interpretation is based on its restriction to \(\smash{x\le L}\). Furthermore, \(\smash{\psi_t\in L^2(\mathbb{R})}\), i.e., it is square integrable over all of $\mathbb{R}$, for any \(\smash{t\ge0}\).  It is important to distinguish the initial momentum-space wave packet $\tps_0$ from the Fourier transform $\mathcal{F}[\psi_0]$ of the full initial solution $\psi_0$, which also contains a reflected component. With that in mind, we will now address \eqref{O1}.

We begin by expressing the detection probability \eqref{PABC} as a spatial integral, using \cite[Eq.\ (8)]{ABC}, \emph{viz.},
\begin{equation}\label{aabove}
    P_{\text{ABC}}\big(\psi_0\big) = \int_{-\infty}^L\!\!\!dx~|\sm\psi_0(x)|^2 \,-\, \lim_{t\to\infty}\int_{-\infty}^L\!\!\!dx~|\sm\psi_t(x)|^2.
\end{equation}
For \(\psi_0(x)\) compactly supported in \((-\infty,L]\) and normalized to unity---
\begin{equation}\label{normal}
    \int_{-\infty}^L\!\!\!dx~|\sm\psi_0(x)|^2= 1 = \int_{-\infty}^{\infty}\!\!dk~|\tps_0(k)|^2,
\end{equation}
\eqref{aabove} can be reformulated as follows:
\begin{equation}\label{abv}
    P_{\text{ABC}}\big(\psi_0\big) = 1 \,-\, \lim_{t\to\infty}\int_0^{\infty}\!\!\!dx~|\sm\psi_t(L-x)\sm|^2.
\end{equation}

Since \(\psi_t(x)\), and consequently \(\psi_t(L-x)\), satisfies Eq.\ \eqref{briefsch} on the full real line, the limit on the right-hand side can be easily evaluated with the help of Dollard's lemma \cite{Dollard}. This lemma states that
\begin{equation}\label{DLemma}
  \lim_{t\to\infty} \int_0^{\infty}\!\!\!dx~\big|\sm\Psi_t(x)\sm\big|^2 = \int_0^{\infty}\!\!\!dk~\big|\sm\tilde{\Psi}_0(k)\sm\big|^2,
\end{equation}
given any solution \(\smash{\Psi_t\in L^2(\mathbb{R})}\) of Schr\"odinger's equation \eqref{briefsch} with initial condition \(\Psi_0(x)\) (or \(\tilde{\Psi}_0(k)=\expval{k\sm|\sm\Psi_0}\) in momentum space).

Since the momentum-space representation of \(\psi_0(L-x)\) is \(e^{i\sm k\sm L}\mathcal{F}[\psi_0](-\sm k)\), it follows from applying \eqref{DLemma} to \eqref{abv} that
\begin{equation}\label{resultABC}
    P_{\text{ABC}}\big(\psi_0\big) = 1 \,-\, \int_0^{\infty}\!\!\!dk~\big|\mathcal{F}[\psi_0](-\sm k)\sm\big|^2.
\end{equation}

To proceed further, note that
\begin{equation}
    \mathcal{F}[\psi_0](k) \overset{\eqref{phitx}}{=} \tps_0(k) \,+\, \rho_\beta(-\sm k)\sm \tps_0(-\sm k)\sm e^{-\,2\sm i\sm k\sm L}.
\end{equation}
Also, incorporating \eqref{normal}, we have
\begin{equation}
    P_{\text{ABC}} = \int_0^{\infty}\!\!\!dk~\left(1-|\sm\rho_\beta(k)|^2\right)\sm|\tps_0(k)\sm|^2 \,-\, \mathcal{R}_L,
\end{equation}
where
\begin{equation}
    \mathcal{R}_L = 2\sm\Re\!\int_0^{\infty}\!\!\!dk~\rho_\beta(k)\sm\tps_0^*(-\sm k)\sm \tps_0(k)\sm e^{2\sm i\sm k\sm L}.
\end{equation}
But this remainder term vanishes in the limit \(\smash{L\to\infty}\), thanks to the Riemann-Lebesgue lemma. Consequently, in this limit, the contrast, Eq.\ \eqref{contrastdef}, reduces to the expression presented in \eqref{contrast}.

This concludes the derivation of \eqref{contrast} for \(\smash{\Re\beta<0}\). However, the same result also obtains for \(\smash{\Re\beta=0}\) applying the \(\epsilon\) prescription outlined below Eq.\ \eqref{rLk}, the details of which we will omit for brevity.

\subsection{~}\label{task2}
By expressing \(\smash{\rho_\beta(k) = 1+2\sm i\sm \beta/(k-i\sm\beta)}\) in \eqref{phitx}, we have
\begin{equation}
    \psi_t(x) = \varphi_t(x) ~+~ \varphi_t(2\sm L-x) ~+~ 2\sm i\sm \beta\,\sm\eta_t(2\sm L-x),
\end{equation}
defining
\begin{subequations}
    \begin{align}
        \varphi_t(x) &:= \int_{-\infty}^{\infty}\!\frac{dk}{\sqrt{2\sm \pi}}~\tps_0(k)\sm e^{i\sm k\sm x-i\sm t\sm k^2/2}, \\
        &\kern-1.9cm\text{and}\nonumber\\
    \eta_t(x) &:= \int_{-\infty}^{\infty}\!\frac{dk}{\sqrt{2\sm \pi}}~\frac{\tps_0(k)}{k-i\sm\beta}\sm e^{i\sm k\sm x-i\sm t\sm k^2/2}.
    \end{align}
\end{subequations}

For the Gaussian initial condition, Eq.\ \eqref{GWP}, these integrals can be easily evaluated. Incorporating the momentum-space wave function \eqref{tpsG}, \(\varphi_t\) simplifies to \eqref{psitG} \cite[Prob.\ 2.22]{grif}, while \(\eta_t\) can be recognized as a representation of the complementary error function \cite[Sec.\ 19.2]{Mfunction}. By combining these elements, we arrive at the final result for \(\psi_t\), which is presented in Eq.\ \eqref{GSol} and referred to as \(\psi_t^{G}\).

It can be verified---either through manual calculations with some effort or more conveniently using software equipped for symbolic calculations---that \(\psi_t^G\) indeed satisfies both the Schr\"odinger equation \eqref{briefsch} and the ABC \eqref{ABCRHS}; in fact, for \emph{any} \(\smash{\beta \in \mathbb{C}}\). However, setting \(\smash{t = 0}\), we reproduce the correct initial condition \eqref{GWP}, along with some small additional terms. These additional terms arise because we derived \(\psi_t^{G}\) using Eq.\ \eqref{phitx}, which only solves the IBVP for initial conditions that are compactly supported in \((-\infty, L]\). In contrast, the Gaussian is supported on all of $\mathbb{R}$.

But since the \(L\) values we are considering are \(\smash{L \gg 1}\), it becomes evident that these additional terms are extremely small. In particular, for \(\smash{\Re\beta < L}\), we can utilize a well-known bound for the complementary error function \cite{erfcbound} to show that for any \(x \in (-\infty, L]\),
\begin{align}
    &\Big|\sm\psi_0^G(x\sm ;k_0,\beta,L) \,-\, e^{i \sm k_0 \sm x}\sm G(x)\sm\Big| \nonumber\\
    &\kern1.2cm < G(L)\left\{1 + \frac{2\sm|\sm\beta\sm|}{L - \Re\beta}\sqrt{1 + \left(\frac{k_0 + \Im\beta}{L - \Re\beta}\right)^2}\,\right\}\!,
\end{align}
which is exponentially small in $L$ due to the $G(L)$ prefactor.

Additionally, for any two solutions of the BVP, \(\psi_t\) and \(\phi_t\), we have
\begin{equation}
    \frac{d}{dt}\sm\big\|\sm\psi_t - \phi_t\sm\big\|^2 = -\sm\Im\beta~\big|\sm\psi_t(L) - \phi_t(L)\sm\big|^2.
\end{equation}
It follows that the squared norm of the difference between any two solutions is a non-increasing function of time (given \(\smash{\Im \beta > 0}\)). This means that the evolution is contractive.

Now, let \(\psi_t\) be the solution given by our derivation \(\big(\psi_t^G\big)\), and let \(\phi_t\) be the solution for the exact Gaussian initial condition restricted to \((-\infty,L]\). We have just shown that the initial difference between them at \(\smash{t=0}\) is exponentially small. Since this difference cannot grow over time, \(\psi_t^G\) remains an excellent approximation to the true solution for all \(\smash{t \ge 0}\), and so these exponentially small deviations from the intended initial conditions can typically be neglected.
\bibliographystyle{unsrt}
\bibliography{ref}
\end{document}